\documentclass[prl,twocolumn,showpacs,showkeys]{revtex4}

\usepackage{amssymb, amsmath, amsfonts, latexsym, verbatim, mathrsfs, graphicx}

\usepackage{graphicx}

\begin{document}

\title{Optimal generation of entanglement under local control}

\author{Raffaele Romano}

\email{rromano@ts.infn.it} \affiliation{The Department of
Theoretical Physics, University of Trieste, Strada Costiera 11,
34014 Trieste, Italy}

\author{Alessio Del Fabbro}

\email{delfabbro@ts.infn.it} \affiliation{The Department of Physics,
University of Trieste, Via Valerio 2, 34127 Trieste, Italy}

%\date{October 31, 2005}

\begin{abstract}

\noindent We study the optimal generation of entanglement between two
qubits subject to local unitary control. With the only assumptions of
linear control and unitary dynamics, by means of a numerical
protocol based on the variational approach (Pontryagin's Minimum
Principle), we evaluate the optimal control strategy leading
to the maximal achievable entanglement in an arbitrary interaction time,
taking into account the energy cost associated to the
controls. In our model we can arbitrarily choose the relative weight
between a large entanglement and a small energy cost.

\end{abstract}



\maketitle

%%%%%%%%%%%%%%%%%%%%%%%%%%%%%%%%%%%%%%%%%%%%%%%%%%%%%%%%%%%%%%%%%%%%%%%%

{\it Introduction.---} The building block for the implementation of
quantum technologies is a pair of interacting two-level systems
(qubits), whose evolution can be affected by external actions. The
basic operations to be accomplished through this system are the
storage and manipulation of encoded information. The use of this
prototype of physical apparatus is motivated by its quantum nature,
that is the existence of some peculiar properties not exhibited by
classical systems. Among them, the quantum correlation called {\it
entanglement} plays a prominent role in the realization of
outperforming protocols.

The standard setting for the manipulation of a pair of qubits is the
so-called {\it Local Unitary control} \cite{khan,benn}. In this
framework, neglecting the influence of the external environment, the
state system is represented by the unit norm complex vector $\vert
\psi (t) \rangle$, satisfying the
Schr\"{o}dinger equation
\begin{equation}\label{eq1}
    \vert \dot{\psi} (t) \rangle = -i \, H_T \bigr( u_1 (t), u_2 (t) \bigl) \vert \psi (t) \rangle,
\end{equation}
where $H_T \bigl( u_1(t), u_2(t) \bigr) = H_1 \bigl( u_1(t) \bigr)
+ H_2 \bigl( u_2(t) \bigr) + H_I$ is the total Hamiltonian.
The local contributions $H_i \bigl( u_i(t) \bigr)$ $(i = 1, 2)$
can be modified by means of external actions, represented by the
control functions $u_1 (t)$ and $u_2 (t)$, and $H_I$ is the
uncontrollable interaction term, responsible for the entanglement
created in the system. We find convenient to express the initial state $\vert \psi_0
\rangle = \vert \psi (0) \rangle$ according to the Schmidt decomposition,
\begin{equation}\label{eq2}
    \vert \psi_0 \rangle = \sqrt{P} \, \vert \varphi \rangle \otimes \vert \chi \rangle +
    \sqrt{1 - P} \, \vert \varphi \rangle^{\bot} \otimes \vert \chi \rangle^{\bot},
\end{equation}
where $P \in [0, 1]$ and $\langle \varphi \vert \varphi \rangle^{\bot} =
\langle \chi \vert \chi \rangle^{\bot} = 0$. If there is not initial correlation between the
two qubits, $P = 0$ or $P = 1$ and $\vert \psi_0 \rangle$ is a product state. Conversely,
a maximally entangled state (Bell state) corresponds to $P = 0.5$.
As a measure of entanglement we introduce the {\it concurrence} $C$ defined as
\begin{equation}\label{eq3}
    C (t) = \vert \langle \psi (t) \vert \sigma_y
    \otimes \sigma_y \vert \psi(t) \rangle^* \vert,
\end{equation}
assuming values in the interval $[0,1]$, vanishing for
uncorrelated states, and reaching its maximum for maximally
entangled states. This quantity satisfies all the properties
of an entanglement monotone \cite{hill}.

Because of it fundamental relevance, several issues regarding the
entanglement generation in the system (\ref{eq1}) have been
addressed in the past years, as well as the problem of generation of
nonlocal gates. In \cite{dur}, D\"{u}r {\it et al.}
characterized the capability of creating entanglement for a generic
interaction $H_I$. In particular, they provided a strategy to
minimize the time of generation of entanglement through arbitrarily
fast local control, and considered the impact of ancillas. Moreover, these
authors proved that an initial amount of entanglement leads to a
more efficient production of entanglement.
In a different context, Kraus and Cirac expressed the maximal attainable
entanglement for an arbitrary unitary operator, as well as the
corresponding initial factorized state \cite{krau}.
The generation of entanglement
implemented via local measurements has been considered in \cite{cira}.

Time-optimality has been further considered in the context of the
simulation of a quantum gate, by introducing the {\it Interaction
Cost}, that is the minimal time to perform a gate using local
operations \cite{vida,hamm,hase}.

While time-optimal procedures are fundamental for the implementation
of efficient computational architectures, they usually ask for impulsive
controls (instantaneous local manipulations of arbitrary strength),
and their operational cost (the energy loss associated to the local
controls) is not accounted for. Moreover, for some physical
apparatus it could be difficult to precisely set the optimal
interaction time, and a predetermined time could be preferable.

With these motivations in mind, in this letter we describe a control
theoretical approach accounting for the aforementioned cost and
considering an arbitrary interaction time. Under the assumption of linear
control with $H_i \bigl( u_i (t) \bigr) = u_i (t) \, H_i$, $i = 1, 2$, and using a
numerical protocol based on the variational method, we evaluate the
best control strategy, that is the optimal control functions $u_i
(t)$ driving an arbitrary initial state $\vert \psi_0 \rangle$ as
close as possible to a maximally entangled state. This is not the most
general form for linear control, however it highly reduces the computational
complexity of the problem while preserving most of its relevant features.
This approach represents a novelty with respect to previous treatments and it
complements them. We will mainly refer to \cite{dur} for a
comparison.

Optimal control methods have been initially used in quantum mechanics
for the control of molecular dynamics \cite{peir}.
Recent applications of these techniques in quantum information
are in particular aimed to determine the optimal gate generation \cite{grac1,grac2,schu}
and the optimal evolution and state transfer (e.g. see \cite{grig,serb}).

%%%%%%%%%%%%%%%%%%%%%%%%%%%%%%%%%%%%%%%%%%%%%%%%%%%%%%%%%%%%%%%%%%%%%%%%

{\it The computational procedure.---} We find the
optimal control strategies $u_i (t)$ using an iterative procedure
based on the variational approach known as {\it Pontryagin Minimum
Principle}. The computational tools employed to derive the results
presented in this work can be used as well for the solution
of more general optimal control problems. In fact, different
performance measures as well as more complicate system dynamics
(for example in the presence of irreversibility and dissipation) can be
imposed. A detailed discussion of the protocol, of its
performance, and of further applications is out of the scope of this
letter, and it will be presented in a forthcoming paper. However,
for sake of completeness, we summarize here the basic ideas underlying
the procedure, without entering into details.

Consider a system described by the state $\textbf{x}(t): \mathbb{R}
\rightarrow {\mathbb R}^n$ whose dynamics and initial conditions are
given by
\begin{equation}\label{eq4}
    \dot{\textbf{x}}(t) = \textbf{f}\bigl( \textbf{x}(t), \textbf{u}(t)
    \bigr), \quad \textbf{x} (0) = \textbf{x}_0,
\end{equation}
where $\textbf{u} (t): \mathbb{R} \rightarrow \mathbb{R}^m$ is a
vector of control functions and $\textbf{f} (\textbf{x},
\textbf{u}): \mathbb{R}^n \times \mathbb{R}^m \rightarrow
\mathbb{R}^n$ is a vector field. Assume that $\textbf{u} (t)$ has to
be chosen such that the cost functional (or performance measure)
\begin{equation}\label{eq5}
    J \bigl( \textbf{u} (t) \bigr) =
    \Phi \bigl( \textbf{x} (\tau) \bigr) + \alpha \int_0^{\tau} {\mathscr L}
    \bigl( \textbf{x} (t), \textbf{u} (t) \bigr) \, dt
\end{equation}
is minimal, where $\tau$ is the fixed final time and $\Phi
(\textbf{x}): \mathbb{R}^n \rightarrow \mathbb{R}$, ${\mathscr L} (\textbf{x},
\textbf{u}): \mathbb{R}^n \times \mathbb{R}^m \rightarrow
\mathbb{R}$ are arbitrary functions. Notice that the cost $J$
contains a final-time term and an integral contribution (that we shall
later denote by $I (\tau)$), and the
real coefficient $\alpha$ expresses the relative weight of them.
Define the {\it optimal control Hamiltonian} as
\begin{equation}\label{eq6}
    {\mathscr H} \bigl( \textbf{x} (t), \textbf{p} (t), \textbf{u} (t)
    \bigr) = \textbf{f} \bigl( \textbf{x} (t), \textbf{u} (t)
    \bigr) \cdot \textbf{p} (t) + {\mathscr L} \bigl( \textbf{x} (t), \textbf{u} (t) \bigr),
\end{equation}
where $\cdot$ is the inner product, and $\textbf{p} (t): \mathbb{R} \rightarrow \mathbb{R}^n$ is an
auxiliary variable conjugate to $\textbf{x} (t)$, often called
costate, whose dynamics is
\begin{equation}\label{eq7}
    \dot{\textbf{p}} (t) = - \nabla_{\textbf{x}} {\mathscr H}
    \bigl( \textbf{x} (t), \textbf{p} (t), \textbf{u} (t) \bigr)
\end{equation}
with final condition
\begin{equation}\label{eq8}
    \textbf{p} (\tau) = - \nabla_{\textbf{x}}
    \Phi \bigl( \textbf{x} (\tau) \bigr).
\end{equation}
Then the Pontryagin Minimum Principle can be stated as follows.

\textbf{Proposition} If we denote by $\textbf{u}^{\prime} (t)$ the
optimal control strategy, and by $\textbf{x}^{\prime} (t)$ and
$\textbf{p}^{\prime} (t)$ the corresponding optimal state and
costate trajectories, then, for all $t \in [0, \tau]$,
\begin{equation}\label{eq9}
    {\mathscr H} \bigl( \textbf{x}^{\prime} (t), \textbf{p}^{\prime} (t),
    \textbf{u}^{\prime} (t) \bigr) \leqslant {\mathscr H} \bigl( \textbf{x}^{\prime} (t),
    \textbf{p}^{\prime} (t), \textbf{u} (t) \bigr).
\end{equation}
Consequently,
\begin{equation}\label{eq10}
    \nabla_{\textbf{u}} {\mathscr H} \bigl( \textbf{x}^{\prime} (t),
\textbf{p}^{\prime} (t), \textbf{u}^{\prime} (t) \bigr) = 0.
\end{equation}

While a more general formulation can be given to this principle (in
particular a non-fixed final time $\tau$ can be considered), this approach
is all we need for our purposes. For more details and for the proof of the
principle, see \cite{kirk} and \cite{dale} (mainly focusing on quantum mechanical
applications).

In our case, we find convenient to represent the state vector $\vert \psi \rangle$
in the computational basis $\{ \vert e_i (t) \rangle, i = 1, \ldots, 4\}$ given by
tensor products of eigenvectors of $\sigma_z$,
\begin{equation}\label{eq11}
    \vert \psi (t) \rangle = \sum_{i = 1}^4 \psi_i (t) \vert e_i \rangle, \quad \psi_i (t) =
    \langle e_i \vert \psi (t) \rangle.
\end{equation}
and $\textbf{x} (t) = \bigl( {\rm {\mathbb R}e} \, \psi_i (t), {\rm
{\mathbb I}m} \, \psi_i (t); i = 1, \ldots, 4 \bigr)^T$, where $T$ means
transposition. Moreover, $\textbf{u} (t) = \bigl( u_1 (t),
u_2 (t) \bigr)^T$, and the vector field $\textbf{f}$ is linear in both $\textbf{x}$ and
$\textbf{u}$. The final-time contribution to the cost function is the deviation of
the final entanglement from its maximal attainable value,
\begin{equation}\label{eq12}
\Phi \bigl( \textbf{x} (\tau) \bigr) = 1 - C (\tau)
\end{equation}
whereas the integral part measures the energy loss, assumed to be proportional
to the squared norm of $\textbf{u}$,
\begin{equation}\label{eq13}
    {\mathscr L} \bigl(\textbf{u} (t)\bigr) = u_1^2 (t) + u_2^2 (t),
\end{equation}
the prototype of energy cost for a nuclear spin driven by a magnetic field.

Our protocol consists of an iteration in which, after solving (\ref{eq4}) and
(\ref{eq7}) with boundary condition (\ref{eq8}), the controls are redefined
step by step in order to fit the condition (\ref{eq10}). The procedure starts
with arbitrary trial functions $u_i (t)$ and it stops when a predetermined
accuracy level is reached.

%%%%%%%%%%%%%%%%%%%%%%%%%%%%%%%%%%%%%%%%%%%%%%%%%%%%%%%%%%%%%%%%%%%%%%%%

{\it Discussion of numerical results.---} The optimal time strategy described in \cite{dur}
is based on the maximization of the entanglement rate at every time. This procedure leads to
a vector $\vert \psi (t) \rangle$ whose (time-dependent) Schmidt coefficient is given by
\begin{equation}\label{eq14}
    P (t) = \sin^2{( h_{max} t + \phi_0 )},
\end{equation}
where $P(0) = P = \sin^2{\phi_0}$. The {\it entanglement capability} $h_{max}$ measures
the ability of the interaction to produce entanglement, its definition and expression in
terms of the singular values of $H_I$ are given in \cite{dur}. The optimal time $\tau_{opt}$ is
defined as the smallest time such that $P (t) = 0.5$. Equation (\ref{eq14}) defines
the steepest entanglement growth, therefore it represents the upper limit for the
production of entanglement up to the optimal time.

Our protocol works for arbitrary Hamiltonian terms and initial states.
In order to illustrate the main results of this work we need to fix them. Because of its
relevance, we consider the Heisenberg interaction $H_I = \sigma_x \otimes \sigma_x
+ \sigma_y \otimes \sigma_y + \sigma_z \otimes \sigma_z$ (with $h_{max} = 2$ and
$\tau_{opt} = \pi / 8$), $H_1 = \sigma_x \otimes {\mathbb I}$, and $H_2 = {\mathbb I} \otimes \sigma_x$.

\begin{figure}[t]
\begin{center} % Requires \usepackage{graphicx}
  \includegraphics[width=9cm]{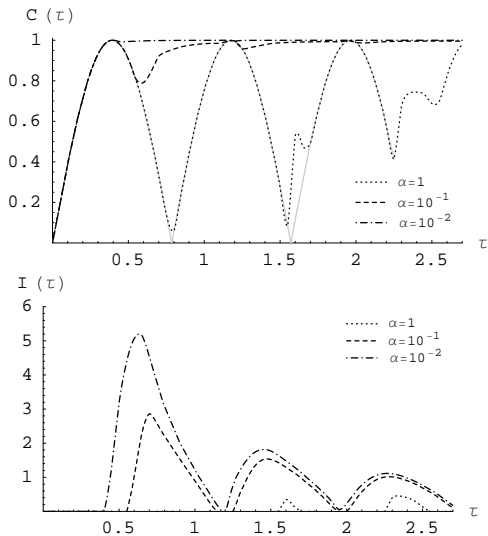} \\
 \caption{\footnotesize Dependence on $\tau$ of the final entanglement $C (\tau)$ and the energy cost
 $I (\tau)$ for several values of the weight parameter $\alpha$. The initial state is an optimal
 state for the Heisenberg Hamiltonian, with $P = 0$, $\vert \varphi \rangle = \vert \uparrow \rangle_z$,
 and $\vert \chi \rangle = \vert \downarrow \rangle_z$.}\label{fig1}
\end{center}
\end{figure}

\begin{figure}[t]
\begin{center} % Requires \usepackage{graphicx}
  \includegraphics[width=9cm]{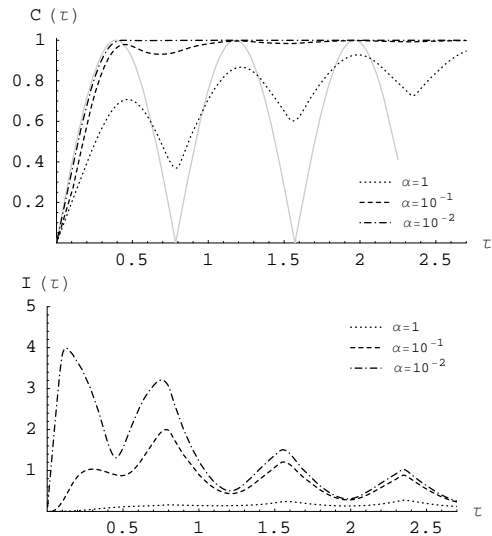} \\
 \caption{\footnotesize Dependence on $\tau$ of the final entanglement $C (\tau)$ and the energy cost
 $I (\tau)$ for several values of the merit function $\alpha$. The initial state is given by
 $P = 0$, $\vert \varphi \rangle = \vert \downarrow \rangle_y$,
 and $\vert \chi \rangle = \vert \downarrow \rangle_z$.}\label{fig2}
\end{center}
\end{figure}

In Fig. \ref{fig1} and \ref{fig2} the dependence on $\tau$ of the final entanglement $C (\tau)$
and the energy cost $I (\tau)$ (the integral in (\ref{eq5})) is shown for several values of
$\alpha$, for the optimal strategies.
The two figures correspond to different uncorrelated initial states ($P = 0$):
$\vert \varphi \rangle = \vert \uparrow \rangle_z$,
$\vert \chi \rangle = \vert \downarrow \rangle_z$ and $\vert \varphi \rangle =
\vert \downarrow \rangle_y$, $\vert \chi \rangle = \vert \downarrow \rangle_z$ respectively
($\vert \uparrow \rangle_i$ and $\vert \downarrow \rangle_i$ denote the $+1$ and $-1$
eigenvectors of the Pauli matrix $\sigma_i$, $i = x, y, z$).

The first state is an optimal initial state for the Heisenberg interaction \cite{dur,krau},
then it evolves in the optimal time to a maximally entangled state without local actions.
This is apparent from Fig. \ref{fig1}: all the represented curves for $C (\tau)$ have the
same behavior for $\tau \leqslant \tau_{opt}$, independently of $\alpha$. These
patterns fit the entanglement evolution associated to (\ref{eq14}), represented by the grey
line, since the system is driven by $H_I$ along the optimal trajectory. Therefore, the
energy cost associated to these paths vanishes. For $\tau > \tau_{opt}$ the parameter $\alpha$
becomes relevant, since a local action is necessary to maximize the final entanglement.
We observe that the goal $C (\tau) = 1$ is approached as $\alpha$ is decreased, that is an
higher cost is tolerated, and that, for every $\alpha$, $C (\tau) \rightarrow 1$
when $\tau$ increases. The peaks of $I (\tau)$ correspond to the valleys associated to
(\ref{eq14}), where a stronger local control is needed, and the magnitude of these peaks
decreases with $\tau$.

\begin{figure}[b]
\begin{center} % Requires \usepackage{graphicx}
  \includegraphics[width=7cm]{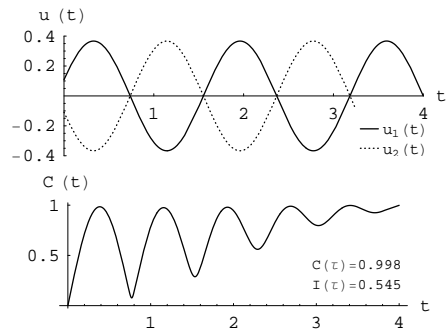} \\
 \caption{\footnotesize Optimal strategies $u_1 (t)$ and $u_2 (t)$, and corresponding
 entanglement $C (t)$, for the optimal initial state, and $\tau = 4$. We have chosen
 $\alpha = 10^{-1}$.}\label{fig3}
\end{center}
\end{figure}

In Fig. \ref{fig2} a similar analysis is presented for a particular non-optimal state. In
this case, local manipulations are needed even for $\tau \leqslant \tau_{opt}$ since the state has
to be adapted in order to fully exploit the entangling capability of the interaction. This is
apparent in the presented plots, where the growth of $C (\tau)$ is steeper with $\alpha$
smaller, and eventually it approaches the optimal curve associated to (\ref{eq14})
Correspondingly, there is a relevant initial contribution to $I (\tau)$. Notice that for an
arbitrary non-optimal state, different patterns could be found for $\alpha$ large,
in particular the steepest growth of $C (\tau)$ could be slower than the optimal one,
and the plateau of $C (\tau)$ could be (even significantly) below $1$. This is due to the
particular choice of $H_1$ and $H_2$, that could be inappropriate for the initial state
considered.

The optimal control strategy can be fixed in line with these considerations.
Given the interaction time $\tau$, the choice of the coefficient $\alpha$
represents a compromise between magnification of $C (\tau)$ and reduction of $I (\tau)$.
The optimal strategies $u_1 (t)$ and $u_2 (t)$ match the requests on $\tau$, $C (\tau)$
and $I (\tau)$. An example is provided in Fig. \ref{fig3}, with the Hamiltonian terms
and the optimal initial state previously introduced, and $\tau = 4$.

Oscillating controls with modulated amplitude are usually obtained when
$\tau$ exceeds the optimal time. When the energy cost is a relevant factor, impulsive
controls are less efficient than controls distributed over time.
They become the optimal strategy when $\tau < \tau_{opt}$, if a large energy expense is accepted.
The function $C (t)$ does not in general stabilize around its maximum (as in the
example presented here), however the amplitude of oscillations usually decreases as $t
\rightarrow \tau$.

\begin{figure}[t]
\begin{center} % Requires \usepackage{graphicx}
  \includegraphics[width=8.5cm]{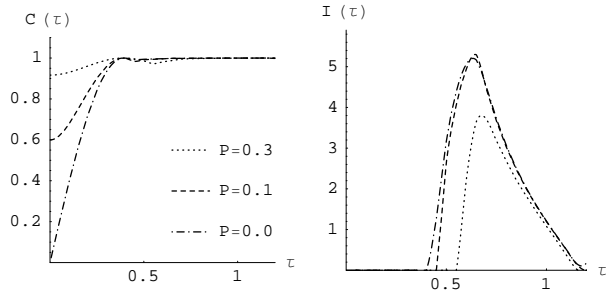} \\
 \caption{\footnotesize Dependence on $\tau$ of the final entanglement $C (\tau)$ and the energy cost
 $I (\tau)$ for several values of the initial Schmidt coefficient $P$. The initial state is defined by
 $\vert \varphi \rangle = \vert \uparrow \rangle_z$, and $\vert \chi \rangle =
 \vert \downarrow \rangle_z$. We have chosen $\alpha = 10^{-2}$.}\label{fig4}
\end{center}
\end{figure}

We have also considered initial states with a non-vanishing entanglement.
Some plots are shown in Fig. \ref{fig4} with $\vert \varphi \rangle =
\vert \uparrow \rangle_z$, $\vert \chi \rangle = \vert \downarrow \rangle_z$
and $P \ne 0$. As intuition suggests, the cost $I (\tau)$ is usually smaller
for correlated initial states.

%%%%%%%%%%%%%%%%%%%%%%%%%%%%%%%%%%%%%%%%%%%%%%%%%%%%%%%%%%%%%%%%%%%%%%%%

{\it Conclusions.---} We have considered a variational approach for the solution of optimal
control problems involving two qubits. By accounting for the energy cost
associated to the manipulations of the system, we are able to find the optimal
strategies to be used in order to drive the system.

In this letter, we have described the entanglement generation for systems
without interaction with the external environment, driven by local
unitary control. Our numerical analysis is consistent with previous results, in particular
it reproduces the optimal entanglement growth and the
corresponding minimal time. Moreover, in our approach the interaction time can be
arbitrarily chosen (for example, it can be a fixed instrumental time or a
predetermined operational time), it is not fixed by the interaction. From this point
of view, our protocol complements the existing methods for the generation of
entanglement, and it is of interest whenever a non-optimal interaction time
is preferred or the energy cost associated to the controls has to be taken
into account.

Using standard Hamiltonian terms and particular initial states, we have studied the relations
among the relevant quantities, and provided some examples. We have found that a
large interaction time is usually preferred for the reduction of the energy
cost, without decreasing the efficiency of the entanglement production. From this
point of view, controls spread over time are more convenient with respect to impulsive
controls.

R. Romano acknowledges support from the European grant
ERG:044941-STOCH-EQ. Work in part supported by INFN, Sezione di
Trieste, Italy.

%%%%%%%%%%%%%%%%%%%%%%%%%%%%%%%%%%%%%%%%%%%%%%%%%%%%%%%%%%%%%%%%%%%%%%%%

\end{document}